\documentclass{article}
\usepackage{spconf,amsmath,graphicx,hyperref}
\usepackage{booktabs}

\ninept

\title{MS-BENCH: A CONDITION-STRATIFIED MULTI-SPEAKER ASR BENCHMARK}

\name{\begin{tabular}{c}
       Chunjiang He$^{1,*}$, Zheng Zi$^{1,*}$, Bingshen Mu$^{1}$, Yurun Li$^{1}$, Chengyou Wang$^{1}$, \\
       Jingyi Wang$^{1}$, Zhonghua Fu$^{1}$, Shuai Wang$^{2}$, Liumeng Xue$^{2}$, Bengu Wu$^{3}$, Hui Bu$^{4}$, \\
       Xin Xu$^{4}$, Zhou Zhu$^{5}$, Qiangze Feng$^{6}$, Daliang Wang$^{6}$, Lei Luo$^{7}$, Qingqing Zhang$^{7}$, Lei Xie$^{1,\dagger}$
      \end{tabular}\thanks{$^*$Equal contribution. $^{\dagger}$Corresponding author.}}

\address{
  $^{1}$Audio, Speech and Language Processing Group (ASLP@NPU), School of Computer Science,\\
       Northwestern Polytechnical University, Xi'an, China\\
  $^{2}$Nanjing University, China\\
  $^{3}$Yutu Zhineng, China\\
  $^{4}$Beijing AISHELL Technology Co., Ltd., Beijing, China\\
  $^{5}$Qualialabs, China\\
  $^{6}$Nexdata Technology Inc., USA\\
  $^{7}$Magic Data Technology Co., Ltd., China}

\begin{document}

%
\maketitle
\begin{abstract}
\vspace{-3.5pt}
Multi-speaker automatic speech recognition (MSASR) has advanced rapidly with end-to-end speech large language models. However, existing benchmarks cover limited ranges of specific multi-speaker conditions and scenarios, making it difficult to reveal system limitations under diverse conditions and scenarios. We introduce MS-Bench\footnote{\href{https://aslp-lab.github.io/MS-Bench/}{Demo page: https://aslp-lab.github.io/MS-Bench/}}, a condition-stratified benchmark for fine-grained MSASR evaluation. MS-Bench integrates both public and internal benchmarks and characterizes each recording along five dimensions: speaker number, overlap ratio, speaker similarity, speaker turn interval, and acoustic quality. We stratify these dimensions into condition ranges and construct MS-Bench with broad coverage of conditions, scenarios, languages, and recording devices. Evaluating representative commercial, open-source, and cascaded systems, we find that overlap affects both lexical recognition and speaker attribution, while other factors exhibit more task-specific effects on recognition, attribution, or temporal prediction. These results demonstrate that aggregate metrics alone obscure important MSASR failure modes and highlight the value of condition-stratified evaluation. MS-Bench aims to reveal condition-specific limitations of current MSASR systems and guide future improvements.

\end{abstract}
\vspace{-4.5pt}
\section{Introduction}  
\label{sec:introduction}
\vspace{-4.5pt}

\begin{figure*}[t]
    \centering
    \includegraphics[width=1.0\textwidth]{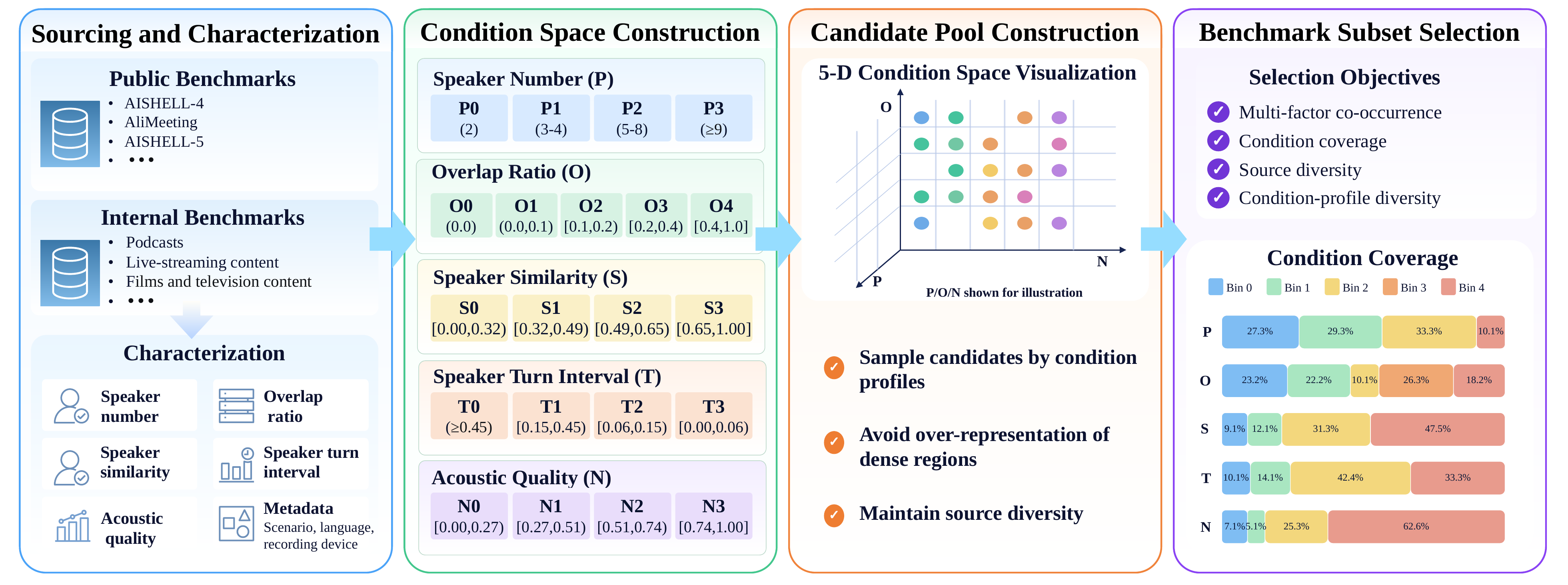}
    \caption{Overview of the MS-Bench construction pipeline.}
    \vspace{-9pt}
    \label{fig:pipeline}
\end{figure*}

Understanding real-world conversations requires accurate recognition of content and a faithful account of how speakers exchange information. Reliable multi-speaker transcription therefore underpins conversational applications, where incorrect speaker attribution can distort the interpretation of otherwise correctly recognized speech. Recognizing speech in the presence of competing speakers is a central aspect of the long-standing cocktail-party problem~\cite{cherry1953}. Beyond recognizing acoustic content, grounding linguistic descriptions in the corresponding audio events and temporal locations has become an important capability for audio understanding models~\cite{xu2021textaudio,wang2026timeaudio,dai2026tagbench}. In multi-speaker conversations, a related requirement is to associate recognized speech with the correct speakers and time intervals. Multi-speaker automatic speech recognition (MSASR) aims to transcribe speech while attributing each segment to its corresponding speaker, thereby addressing the critical question of ``who spoke what and when''. Compared with single-speaker ASR, MSASR remains challenging due to overlapping speech, rapid speaker transitions, speaker similarity, and diverse acoustic conditions in real-world scenarios~\cite{sot,saasr,lin2026speakerreasoner}. Traditional cascaded MSASR pipelines optimize speaker diarization, segmentation, and ASR as separate modules~\cite{mu2026summary,alimeeting,survey}, which can lead to error propagation.
Recent advances in audio and speech large language models have demonstrated strong capabilities in audio understanding and reasoning, including long-form audio processing and temporal perception~\cite{ghosh2025audioflamingo3,xu2025qwen3omni,yang2026mossaudio}. End-to-end MSASR systems based on speech large language models (SLLMs), such as DiaScriber~\cite{diascriber}, VibeVoice-ASR~\cite{vibevoice-asr}, SoulX-Transcriber~\cite{soulx-transcriber}, and MOSS-Transcribe-Diarize~\cite{moss}, have achieved competitive performance on widely used multi-speaker benchmarks, including AISHELL-4~\cite{aishell4}, AliMeeting~\cite{alimeeting}, MagicData-RAMC~\cite{magicdata-ramc}, AMI~\cite{ami}, and MLC-SLM~\cite{mu2026summary}. These advances raise an important question: can current audio LLM-based systems reliably perform MSASR under challenging cocktail-party-like conditions, correctly recognizing speech and grounding it to the corresponding speakers and time intervals? However, existing benchmarks remain insufficient to comprehensively answer this question. First, they generally lack systematic coverage across different levels of key multi-speaker conditions, including speaker number, overlap ratio, speaker similarity, speaker turn interval, and acoustic quality, making it difficult to evaluate MSASR performance across these factors. Moreover, most benchmarks focus on specific application scenarios: AISHELL-4, AliMeeting, and AMI mainly target meeting speech, whereas MagicData-RAMC and MLC-SLM focus on conversational speech. Consequently, these limitations make it difficult to reveal condition-specific weaknesses of current MSASR systems and guide further system improvement.

In this work, we introduce MS-Bench, a condition-stratified benchmark for fine-grained evaluation of MSASR systems. MS-Bench assesses a key aspect of the cocktail-party problem by systematically evaluating how well MSASR systems determine what was spoken, who spoke it, and when it occurred across diverse multi-speaker conditions. MS-Bench integrates multiple public and internal benchmarks and characterizes each recording along five dimensions: speaker number, overlap ratio, speaker similarity, speaker turn interval, and acoustic quality. We stratify these dimensions into different condition ranges and select recordings to achieve broad condition coverage while balancing the remaining factors. In this way, MS-Bench provides broad coverage of multi-speaker conditions, application scenarios, languages, and recording devices, enabling more systematic analysis of system performance under different conditions. We evaluate representative commercial, open-source, and cascaded MSASR systems and analyze their performance at both the system level and across the five condition dimensions. The results show that overlapping speech jointly degrades lexical recognition and speaker attribution; speaker similarity and turn transitions increase speaker-attribution confusion; acoustic degradation primarily affects lexical recognition accuracy and is associated with shorter predicted speech timelines; and higher speaker counts are associated with greater attribution difficulty in most systems after controlling for dataset and overlap. Overall, these findings suggest that the evaluated systems do not yet achieve consistently reliable MSASR under challenging cocktail-party-like conditions, highlighting the need for further improvements in overlap-robust recognition, speaker attribution, and temporal modeling.

\vspace{-4.5pt}
\section{MS-Bench}   
\label{sec:bench}
\vspace{-4.5pt}

\subsection{Sourcing and Characterization}   
\label{sssec:stage1}
\vspace{-4.5pt}

We first collect recordings from publicly available and internal MSASR benchmarks.
The publicly available benchmarks include representative multi-speaker benchmarks, such as AISHELL-4, AliMeeting, and AISHELL-5~\cite{aishell5}. Each recording is characterized from three perspectives: speaker interaction, acoustic condition, and recording metadata. Specifically, we extract the number of speakers, overlap ratio, speaker similarity, and speaker turn intervals from the speaker annotations. We measure speaker similarity by extracting CAM++~\cite{cam++} embeddings from 3--5 non-overlapping segments longer than 2 seconds for each speaker and computing pairwise cosine similarities. We calculate speaker turn intervals only between consecutive non-overlapping speaker turns to avoid the influence of overlapping regions. In addition, we evaluate acoustic quality using DNSMOS P.835~\cite{dnsmos} and NISQA~\cite{nisqa}, and collect metadata information including scenario, language, and recording device.

\vspace{-4.5pt}
\subsection{Condition Space Construction}   
\label{sssec:stage2}
\vspace{-4.5pt}
Based on the recording-level characteristics obtained above, we construct a unified condition space by partitioning five dimensions: speaker number, overlap ratio, speaker similarity, speaker turn interval, and acoustic quality. We group speaker number using predefined count ranges, while we manually partition the overlap ratio by severity to preserve the interpretability of different overlap levels. We divide other continuous attributes, including speaker similarity, speaker turn interval, and acoustic quality, according to empirical quantiles to maintain sufficient sample support within each range. Specifically, we represent speaker similarity by the maximum pairwise cosine similarity within each recording to capture the most confusable speaker pair. Speaker turn interval is represented by its 25th percentile to characterize relatively rapid speaker transitions while reducing the influence of long pauses and extreme values. For acoustic quality, we jointly consider DNSMOS P.835 and NISQA. Because these metrics have different score distributions, we first convert their scores to percentile ranks, denoted $p_D$ and $p_N$, and combine them as $Q=(p_D+p_N)/2$. We further define $d_N=1-Q$, where larger values indicate poorer acoustic quality. Fig.~\ref{fig:pipeline} summarizes the resulting condition ranges.
\vspace{-4.5pt}
\subsection{Candidate Pool Construction}
\label{stage3}
\vspace{-4.5pt}
Using the constructed condition space, we build an intermediate candidate pool to reduce bias from uneven condition distributions across source datasets. Instead of sampling recordings by dataset identity, we identify candidates based on their condition profiles, covering different ranges across the five condition dimensions while avoiding over-representation of densely populated regions. We also consider source diversity to prevent any individual dataset from dominating the candidate pool. The resulting pool provides a structured search space for constructing the final benchmark subset.

\vspace{-4.5pt}
\subsection{Benchmark Subset Selection}   
\label{sssec:stage4}
\vspace{-4.5pt}

Given the structured candidate pool, we construct the final benchmark subset by jointly optimizing condition coverage, multi-factor co-occurrence, condition-profile diversity, and source diversity. To obtain a compact yet representative subset, each candidate recording $i$ is assigned a selection utility $U_i = 0.40C_i + 0.25M_i + 0.20D_i + 0.15R_i$, where $C_i$ measures the contribution to condition coverage, $M_i$ characterizes the co-occurrence of multiple challenging conditions, $D_i$ measures diversity gain in the condition space, and $R_i$ encourages the inclusion of under-represented sources. We obtain the final subset by maximizing total utility under constraints on subset size, minimum coverage over predefined operating ranges, and source distribution. Finally, we manually inspect audio integrity, transcripts, speaker annotations, and timestamps. We replace samples that fail quality control with eligible candidates with similar condition profiles.

\vspace{-9pt}
\subsection{Benchmark Overview}  
\label{ssec:overview}
\vspace{-4.5pt}

MS-Bench is a condition-stratified benchmark with a total duration of 32.72 hours. Recording durations range from 1.38 to 159.07 minutes, with a median of 16.50 minutes. The benchmark covers a broad range of speech scenarios, spanning meetings, spontaneous conversations, films and television content, podcasts, dinner-party conversations, educational content, live-streaming content, in-vehicle interactions, and smart-glasses interactions. The recordings mainly contain Chinese and English speech, with additional samples in Portuguese, Japanese, Thai, Italian, and Spanish. They are collected using heterogeneous recording devices, including microphone arrays, in-cabin recording systems, and mobile devices.
 Beyond scenario and metadata diversity, MS-Bench systematically covers the five condition dimensions defined in Sec.~\ref{sssec:stage2}: speaker number, overlap ratio, speaker similarity, speaker turn interval, and acoustic quality. The number of speakers ranges from 2 to 14, and the duration-weighted average overlap ratio is 21.87\%. As shown in Fig.~\ref{fig:pipeline}, the benchmark covers all predefined operating ranges, enabling condition-stratified analysis of multi-speaker transcription systems under diverse interaction and acoustic conditions.

\label{sec:ex-setup}
\label{ssec:evaluated_systems}

\begin{table*}[t]
  \centering
  \caption{Overall performance and robustness on MS-Bench. DER, cpWER, tcpWER, and WER are reported in \%. ``Excluded'' denotes units omitted from metric computation due to invalid or unparsable outputs, while ``Soft Deg. Units'' (soft degradation units) denotes units with degraded but still parsable outputs. Bold indicates the best performance, underlining indicates the second-best performance.}
  \label{tab:overall_performance}
  \scriptsize
  \setlength{\tabcolsep}{6pt}
  \renewcommand{\arraystretch}{1.0}

  \begin{tabular}{l cccccc}
    \toprule
    \textbf{System} &
    \textbf{DER} &
    \textbf{cpWER} &
    \textbf{tcpWER} &
    \textbf{WER} &
    \textbf{Excluded} &
    \textbf{Soft Deg. Units} \\
    \midrule

    MOSS-Transcribe-Diarize-Pro
    & \textbf{10.96} & \textbf{21.95} & \textbf{22.81} & \textbf{22.49} & 15 & 2 \\

    MOSS-Transcribe-Diarize
    & \underline{12.74} & \underline{24.94} & \underline{26.22} & \underline{25.02} & 0 & 2 \\

    Qwen3.8-Omni-Flash
    & 17.97 & 29.43 & 32.59 & 27.86 & 5 & 2 \\

    VibeVoice-ASR
    & 20.19 & 43.41 & 44.76 & 33.11 & 10 & 12 \\

    pyannote+Qwen3-ASR
    & 17.35 & 43.25 & 46.54 & 34.96 & 0 & 1 \\

    Doubao-ASR 2.0
    & 39.52 & 55.37 & 62.07 & 38.20 & 4 & 26 \\

    SoulX-Transcriber (5-min segments)
    & 19.42 & 49.16 & 52.60 & 39.96 & 8 & 30 \\

    \bottomrule
  \end{tabular}

  \vspace{-9pt}
\end{table*}


\vspace{-4.5pt}
\section{Evaluation Setup and Metrics}
\vspace{-4.5pt}
\label{ssec:evaluation_metrics}   

We evaluate the systems using diarization error rate (DER), concatenated minimum-permutation word error rate (cpWER), time-constrained minimum-permutation word error rate (tcpWER)~\cite{meeteval}, and word error rate (WER). DER measures speaker diarization performance with a 0.5-second collar. cpWER evaluates speaker-attributed transcription by finding the speaker permutation that minimizes the overall word error rate. tcpWER further incorporates temporal constraints into speaker--transcript matching with a 5-second collar. Speaker-agnostic WER evaluates transcription accuracy independently of speaker assignment by disregarding speaker labels during scoring.

An inference unit is the audio span submitted in one independent system request and scored against the corresponding reference span. Recordings are kept intact whenever supported; those exceeding the system-specific input limits are split into chunks of at most 90 minutes for the MOSS systems, 60 minutes for VibeVoice-ASR, and 5 minutes for SoulX-Transcriber, while the remaining systems process the original recordings without chunking. We score independently inferred chunks against their corresponding reference windows because speaker IDs need not remain consistent across requests. Units are excluded only when their outputs cannot be reliably scored, including API refusal, structurally incomplete generation at the token limit, invalid intervals with $\mathrm{end}<\mathrm{start}$, or a language mismatch with the reference. Parseable repetitive or collapsed outputs remain in scoring and are reported as soft degradation.


For diagnostic analysis, we decompose the standard WER numerator
$\mathrm{S}+\mathrm{I}+\mathrm{D}$ into two mutually exclusive sources.
Substitutions are treated as lexical errors, while every insertion and
deletion is assigned to either lexical recognition or segmentation
according to its cause. Specifically, \(E_{\mathrm{lex}} = (\mathrm{S}+\mathrm{I}_{\mathrm{lex}}+\mathrm{D}_{\mathrm{lex}})/N_{\mathrm{ref}}, E_{\mathrm{seg}} = (\mathrm{I}_{\mathrm{seg}}+\mathrm{D}_{\mathrm{seg}})/N_{\mathrm{ref}}\),
where $\mathrm{I}_{\mathrm{seg}}$ and $\mathrm{D}_{\mathrm{seg}}$
denote content introduced or lost due to falsely detected, missed, or
discarded speech regions. Thus,
$E_{\mathrm{lex}}+E_{\mathrm{seg}}=\mathrm{WER}$.
Speaker attribution is evaluated separately by
$E_{\mathrm{attr}}$, the fraction of reference utterances attributed
to an incorrect speaker.



For condition-wise comparisons, $\Delta_x$ denotes the difference in error rate between the two conditions being compared, in percentage points, where $x\in{\mathrm{lex},\mathrm{seg},\mathrm{attr}}$. We define the two conditions separately for each analysis, and restrict comparisons to inference units containing samples from both conditions, reducing between-unit confounding. We estimate 95\% confidence intervals by resampling recordings with replacement and recomputing $\Delta_x$; this recording-level cluster bootstrap keeps all utterances and chunks from the same recording together during resampling.

\begin{table*}[!t]
  \centering
    \caption{Condition-wise diagnostic results across the seven evaluated systems. Entries report paired error-rate differences or Spearman correlations, depending on the corresponding condition analysis. $^{*}$ indicates statistical significance as defined in Sec.~\ref{ssec:evaluation_metrics}.}
  \label{tab:condition_analysis}
  \scriptsize
  \setlength{\tabcolsep}{2.7pt}
  \renewcommand{\arraystretch}{0.96}
  \begin{tabular}{l cc c ccc c c}
    \toprule
    & \multicolumn{2}{c}{\textbf{Overlap}}
    & \textbf{Speaker similarity}
    & \multicolumn{3}{c}{\textbf{Acoustic quality}}
    & \textbf{Speaker turn interval}
    & \textbf{Speaker number} \\
    \cmidrule(lr){2-3}\cmidrule(lr){4-4}\cmidrule(lr){5-7}\cmidrule(lr){8-8}\cmidrule(lr){9-9}
    \textbf{System}
    & $\Delta_{\mathrm{lex}}$ & $\Delta_{\mathrm{attr}}$
    & $\rho_{\mathrm{sim}}$
    & $\rho_{\mathrm{lex}}$ & $\rho_{\mathrm{attr}}$ & $\rho_{\mathrm{time}}$
    & $\Delta_{\mathrm{attr}}$
    & $\rho_{\mathrm{attr}\mid D,O}$ \\
    \midrule
    MOSS-Transcribe-Diarize-Pro & +27.3$^{*}$ & +10.6$^{*}$ & +0.311$^{*}$ & +0.392$^{*}$ & $-$0.060 & $-$0.350$^{*}$ & +2.49$^{*}$ & +0.403$^{*}$ \\
    MOSS-Transcribe-Diarize & +27.3$^{*}$ & +10.9$^{*}$ & +0.312$^{*}$ & +0.480$^{*}$ & $-$0.121 & $-$0.405$^{*}$ & +2.48$^{*}$ & +0.350$^{*}$ \\
    Qwen3.8-Omni-Flash & +28.7$^{*}$ & +10.5$^{*}$ & +0.323$^{*}$ & +0.364$^{*}$ & $-$0.063 & $-$0.335$^{*}$ & +2.74$^{*}$ & +0.257 \\
    VibeVoice-ASR & +34.3$^{*}$ & +4.2$^{*}$ & +0.297$^{*}$ & +0.569$^{*}$ & +0.081 & $-$0.436$^{*}$ & +3.07$^{*}$ & +0.054 \\
    pyannote+Qwen3-ASR & +38.6$^{*}$ & +19.1$^{*}$ & +0.248$^{*}$ & +0.392$^{*}$ & +0.030 & $-$0.437$^{*}$ & +3.43$^{*}$ & +0.438$^{*}$ \\
    Doubao-ASR 2.0 & +27.2$^{*}$ & +6.3$^{*}$ & +0.268$^{*}$ & +0.304$^{*}$ & $-$0.057 & $-$0.145 & +4.17$^{*}$ & +0.281$^{*}$ \\
    SoulX-Transcriber (5-min segments) & +35.2$^{*}$ & +8.7$^{*}$ & +0.319$^{*}$ & +0.181$^{*}$ & $-$0.174$^{*}$ & $-$0.246$^{*}$ & +1.51$^{*}$ & -- \\
    \bottomrule
  \end{tabular}
  \vspace{-9pt}
\end{table*}

\vspace{-4.5pt}
\section{RESULTS AND ANALYSIS}   
\label{sec:results}
\vspace{-4.5pt}
We evaluate a cascaded MSASR system, end-to-end MSASR models, commercial ASR services, and omni-modal LLMs. Specifically, we evaluate MOSS-Transcribe-Diarize-Pro\footnote{\href{https://platform.mosi.cn/docs/scenarios/transcribe-diarization}{MOSS-Transcribe-Diarize-Pro webpage}}~\cite{moss}, MOSS-Transcribe-Diarize~\cite{moss}, VibeVoice-ASR~\cite{vibevoice-asr}, SoulX-Transcriber~\cite{soulx-transcriber}, Doubao ASR 2.0\footnote{\href{https://console.volcengine.com/ark/model/detail?name=doubao-seed-asr-2-0}{Doubao-ASR 2.0 webpage}}, Qwen3.8-Omni-Flash\footnote{\href{https://modelstudio.console.alibabacloud.com/ap-southeast-1/model/market/detail/qwen3.8-omni-flash}{Qwen3.8-Omni-Flash webpage}}~\cite{qwen3.8-omini}, and a cascaded system using pyannote speaker-diarization-community-1~\cite{pyannote-audio1, pyannote-audio2} followed by Qwen3-ASR~\cite{qwen3-asr}.

\vspace{-4.5pt}
\subsection{Overall Performance}
\label{ssec:overall_performance}
\vspace{-4.5pt}

Table~\ref{tab:overall_performance} summarizes the overall performance and robustness of the evaluated systems. MOSS-Transcribe-Diarize-Pro achieves the best overall performance, with the lowest DER (10.96\%), cpWER (21.95\%), tcpWER (22.81\%), and WER (22.49\%), followed by MOSS-Transcribe-Diarize. Qwen3.8-Omni-Flash achieves the next-lowest cpWER (29.43\%) and WER (27.86\%), while the cascaded pyannote+Qwen3-ASR system obtains a similar DER (17.35\% vs.\ 17.97\%) but higher cpWER (43.25\% vs.\ 29.43\%).

The systems also differ in output robustness. MOSS-Transcribe-Diarize and the cascaded system have no excluded units, whereas MOSS-Transcribe-Diarize-Pro excludes 15 units: three are rejected by the API, and 12 contain invalid timestamps. Thus, although MOSS-Transcribe-Diarize-Pro achieves lower error rates on the valid outputs, MOSS-Transcribe-Diarize provides better output robustness and coverage. Qwen3.8-Omni-Flash has five excluded and two soft-degradation units, further showing that accuracy and robustness do not always vary consistently across systems.

\begin{figure}[t]
\centering
\includegraphics[width=\columnwidth]{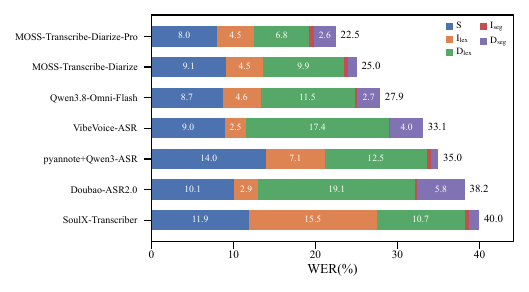}
\vspace{-26pt}
\caption{Error composition by processing stage, the first three segments form $E_{\mathrm{lex}}$ and the last two $E_{\mathrm{seg}}$, whose sum is WER (labelled at the right of each bar). Systems are sorted by WER, best first; segments smaller than 2~pp are not labelled.}
\vspace{-14pt}
\label{fig:composition}
\end{figure}

Figure~\ref{fig:composition} decomposes WER into $E_{\mathrm{lex}}$ and $E_{\mathrm{seg}}$: lexical errors dominate for every system, with $E_{\mathrm{lex}}$ spanning 19.27\%--38.17\% whereas $E_{\mathrm{seg}}$ stays within 1.32\%--6.02\%. The composition is nevertheless system-specific---MOSS-Transcribe-Diarize-Pro gains over MOSS-Transcribe-Diarize mainly by lowering $E_{\mathrm{lex}}$ (23.48\%$\rightarrow$19.27\%) at the cost of a higher $E_{\mathrm{seg}}$ (1.54\%$\rightarrow$3.22\%), while SoulX-Transcriber is dominated by insertions ($\mathrm{I}_{\mathrm{lex}}$ 15.5\%) rather than by recognition or segmentation loss. Error mechanisms therefore differ across systems even when aggregate WER is close, which motivates the condition-wise decomposition in Sec.~\ref{ssec:condition_analysis}.

\vspace{-4.5pt}
\subsection{Condition-stratified Analysis}
\label{ssec:condition_analysis}
\vspace{-4.5pt}
\textbf{Overlap.} 
For each reference utterance $u$, $r_{\mathrm{ov}}(u)$ is the fraction of its duration covered by the speech intervals of other speakers.
We compare only $r_{\mathrm{ov}}=0$ and $r_{\mathrm{ov}}\geq0.5$, and retain inference units containing both conditions.
This within-inference-unit comparison keeps the recording source, acoustic environment, and system input context fixed, thereby reducing between-recording confounding.
Table~\ref{tab:condition_analysis} shows that heavy overlap significantly increases $E_{\mathrm{lex}}$ by 27.2--38.6 pp and $E_{\mathrm{attr}}$ by 4.2--19.1 pp across all seven systems, indicating that overlap consistently degrades both lexical recognition and speaker attribution.
To further distinguish missed speech from overlap-localization and speaker-assignment errors, we examine the system outputs within reference overlap regions. We first evaluate whether the system temporally captures the onset and offset of an overlap region: a reference overlap boundary is considered localized if a hypothesized speech boundary occurs within $\pm0.5$ s of it. Under this criterion, only 39.9--85.3\% of reference overlap boundaries are successfully localized across systems, indicating substantial variation in their ability to represent the temporal structure of overlapping speech. Although 71--93\% of overlapped speech is covered by at least one hypothesized speaker, only 31--63\% is covered by the correct speaker. These results show that overlap-related failures are not simply caused by missing speech altogether: systems often detect speech activity in overlap regions, but may fail to represent the overlap boundaries accurately and, more importantly, assign a substantial portion of the detected speech to the wrong speaker.


\textbf{Speaker similarity.}
For an ordered pair of true speakers $(s_i,s_j)$, we define the pairwise confusion $C_{i\rightarrow j}$ as the fraction of aligned tokens from $s_i$ that the system attributes to $s_j$, and relate it to the speaker similarity $\mathrm{Sim}_{ij}$ of the two speakers. When the analysis is restricted to completely non-overlapping utterances, Table~\ref{tab:condition_analysis} shows significant positive correlations between $\mathrm{Sim}_{ij}$ and $C_{i\rightarrow j}$ for all seven systems, with Spearman's $\rho$ ranging from 0.248 to 0.323. Thus, the more acoustically similar two speakers are, the more likely content from one speaker is to be attributed to the other, indicating that acoustic speaker similarity is a consistent source of attribution error even in the absence of overlap.



\textbf{Acoustic quality.}
%
To separate the effect of acoustic quality from overlap and speaker-transition effects, we restrict this analysis to utterances without overlap and away from speaker-turn boundaries, and correlate the recording-level acoustic degradation score $d_N$ with each error measure. Table~\ref{tab:condition_analysis} shows significant positive correlations between $d_N$ and $E_{\mathrm{lex}}$ for all seven systems, with Spearman's $\rho$ ranging from 0.181 to 0.569. In contrast, the correlation with $E_{\mathrm{attr}}$ is weak and significant for only one system. These results indicate that acoustic degradation primarily affects lexical recognition rather than speaker attribution.
We further examine whether acoustic degradation changes the temporal extent of predicted speech. Let
$
R_{\mathrm{time}}={T_{\mathrm{hyp}}}/{T_{\mathrm{ref}}},
$
where $T_{\mathrm{hyp}}$ and $T_{\mathrm{ref}}$ denote the total hypothesized and reference speech durations, respectively. In Table~\ref{tab:condition_analysis}, $\rho_{\mathrm{time}}$ denotes the Spearman correlation $\rho(d_N,R_{\mathrm{time}})$. The table shows significantly negative correlations for six of the seven systems, indicating that poorer acoustic conditions are generally associated with shorter predicted speech timelines. We interpret this result specifically as systematic temporal contraction rather than timestamp-boundary error, since $R_{\mathrm{time}}$ measures the overall temporal extent of hypothesized speech and is insensitive to a uniform temporal shift.

\textbf{Speaker turn interval.}
We identify non-overlapping reference speaker switches, expand each boundary by $\pm\delta$, and, within the same inference unit, compare utterances intersecting a switch neighborhood with utterances far from all switches. This controls recording factors while testing the local effect of a turn boundary. Table~\ref{tab:condition_analysis} reports $\delta=0.5$~s, for which $E_{\mathrm{attr}}$ increases significantly by 1.51--4.17~pp in all systems. Neither $E_{\mathrm{lex}}$ nor $E_{\mathrm{seg}}$ is significant in any system, and the same qualitative pattern holds for $\delta\in\{0.25,1.0\}$~. Speaker transitions therefore primarily disrupt \emph{who spoke}, rather than \emph{what was said}.

\textbf{Speaker number.}
Speaker number is strongly coupled with dataset composition in our benchmark: in particular, all recordings in the $P3$ ($\geq9$ speakers) group come from one dataset. We therefore avoid interpreting the pooled $P0$--$P3$ trend as a monotonic effect of speaker count. For the controlled analysis, we exclude $P3$ to reduce this dataset-specific bias, and exclude SoulX-Transcriber because its 5-min inference units do not always contain all speakers present in the original recording.
To reduce dataset confounding, we compare recordings relative to others from the same dataset rather than directly across datasets. Specifically, we center both speaker count and $E_{\mathrm{attr}}$ by subtracting their respective dataset means, and then compute the Spearman correlation between the two residualized variables. Thus, the analysis asks whether a recording with more speakers than the dataset average also tends to exhibit higher attribution error than the dataset average. In a stricter analysis, we additionally remove the effect of overlap ratio from both variables before computing the correlation. We denote this correlation as $\rho_{\mathrm{attr}\mid D,O}$, where $D$ and $O$ denote dataset and overlap ratio, respectively. Table~\ref{tab:condition_analysis} shows positive associations for all six evaluated systems; five are significant after controlling for dataset, and four remain significant after additionally controlling for overlap. These results indicate that higher speaker count is associated with greater speaker-attribution difficulty beyond dataset composition and, for most systems, beyond overlap alone, without implying causality.

\vspace{-4.5pt}
\section{CONCLUSION}  
\vspace{-4.5pt}

This work introduces MS-Bench, a condition-stratified benchmark for systematic evaluation of multi-speaker transcription systems. By characterizing recordings along five dimensions covering speaker interaction, speaker similarity, and acoustic conditions, MS-Bench provides a unified framework for analyzing system performance beyond dataset-level evaluation. Experiments on representative commercial, open-source, and cascaded systems reveal that overlapping speech degrades both lexical recognition and speaker attribution, while speaker similarity and rapid turn transitions increase attribution confusion. Acoustic degradation is further associated with increased lexical errors and shorter predicted speech timelines. These findings indicate that, despite recent advances in audio and speech LLMs, reliable MSASR under challenging cocktail-party-like conditions remains an open problem for the evaluated systems. MS-Bench highlights overlap-robust recognition, accurate speaker attribution, and faithful temporal modeling as important directions for future research.
\bibliographystyle{IEEEbib}
\bibliography{strings,refs}

\end{document}